\documentclass[12pt,preprint]{aastex}

\newcommand{\gtsim}{\protect\raisebox{-0.5ex}{$\:\stackrel{\textstyle >}{\sim}\:$}}

\slugcomment{Submitted to ApJ}

\shorttitle{ Neutrino flux predictions for known Galactic microquasars }
\shortauthors{Distefano et al.}

\begin{document}

\title{ Neutrino flux predictions for known Galactic microquasars }

\author{C. Distefano\altaffilmark{1,2}, D. Guetta \altaffilmark{3}, E. Waxman \altaffilmark{4} and  A. Levinson\altaffilmark{5} }

\altaffiltext{1}{ Dipartimento di Fisica e Astronomia,
Universit\'a di Catania, Catania. }
\altaffiltext{2}{ Laboratori Nazionali del Sud, Catania. }
\altaffiltext{3}{ Osservatorio di Arcetri, L. E. Fermi 2, Firenze. }
\altaffiltext{4}{ Department of Condensed Matter Physics, Weizmann
Institute, Rehovot 76100, Israel. }
\altaffiltext{5}{ School of Physics and Astronomy, Tel Aviv
University, Tel Aviv 69978, Israel }

\begin{abstract}
It has been proposed recently that Galactic microquasars may be
prodigious emitters of TeV neutrinos that can be detected by
upcoming km$^2$ neutrino telescopes. In this paper we consider a
sample of identified microquasars and microquasar candidates, for
which available data enables rough determination of the jet
parameters. By employing the parameters inferred from radio
observations of various jet ejection events, we determine the
neutrino fluxes that should have been produced during these events
by photopion production in the jet. Despite the large
uncertainties in our analysis, we demonstrate that in several of
the sources considered, the neutrino flux at Earth, produced in
events similar to those observed, would exceed the detection
threshold of a km$^2$ neutrino detector. The class of microquasars
may contain also sources with bulk Lorentz factors larger than
those characteristic of the sample considered here, directed along
our line of sight. Such sources, which may be very difficult to
resolve at radio wavelengths and hence may be difficult to
identify as microquasar candidates, may emit neutrinos with fluxes
significantly larger than typically obtained in the present
analysis. These sources may eventually be identified through their
neutrino and gamma-ray emission.
\end{abstract}


\section{Introduction}

Microquasars are Galactic X-ray binary (XRB) systems, which
exhibit relativistic radio jets \cite{Mirabel99,Fe01}. These
systems are believed to consist of a compact object, a neutron
star or a black hole, and a giant star companion. Mass transfer
from the giant star to the compact object through the formation of
an accretion disc and the presence of the jets make them similar
to small quasars, hence their name ``microquasars.'' The analogy
may not be only morphological; although there is no obvious
scaling, it is common thinking that the physical processes that
govern the formation of the accretion disc and the ejection of
plasma into the jets are the same for both systems. Local,
Galactic microquasars may therefore be considered as nearby
``laboratories,'' where models of the distant, powerful quasars
can be tested. Microquasars are associated with several classes of
XRBs, and differ in their time behaviour. Some, like SS433, are
persistent sources, while others appear to be intermittent (GRS
1915+105) or periodic (LS I +61$^\circ$303). In all cases,
however, the observed radiation from microquasar jets, typically
in the radio and in some cases also in the IR band, is consistent
with non-thermal synchrotron radiation emitted by a population of
relativistic, shock-accelerated electrons.

The composition of microquasar jets is still an open issue. The
synchrotron emission both in the radio and in the IR is consistent
with near equipartition between electrons and magnetic field,
which is also implied by minimum energy considerations
(Levinson \& Blandford 1996). However, the dominant energy carrier in the
jet is presently unknown (with the exception of the jet in SS433).
Scenarios whereby energy extraction is associated with spin down
of a Kerr black hole favor $e^{\pm}$ composition (although baryon
entrainment is an issue). However, the pair annihilation rate
inferred from the estimated jet power implies electromagnetic
domination on scales smaller than roughly 10$^9$~cm in the
superluminal sources, and requires a transition from
electromagnetic to particle dominated flow above the annihilation
radius by some unspecified mechanism \cite{Levinson96}.
Alternatively, in scenarios in which an initial rise of the X-ray
flux leads to ejection of the inner part of the accretion disc, as
widely claimed to be suggested by the anti-correlation between the
X-ray and radio flares seen during major ejection events, e-p jets
are expected to be produced. A possible diagnostic of e-p jets is
the presence of Doppler-shifted spectral lines such as the
H$_{\alpha}$ line observed in SS433, proving the presence of
protons in the jets of this source. Unfortunately, detection of
such  ``smoking gun'' lines from jets having Lorentz factors well
in excess of unity (as is the case in the superluminal
microquasars) may  be far more difficult than in SS433, as the
lines are anticipated to be very broad ($\Delta
\lambda/\lambda\gtsim$ 0.1).  Furthermore, the conditions required
to produce detectable flux in such sources may be far more extreme
than in SS433. Another diagnostic of hadronic jets, namely the
emission of TeV neutrinos, has recently been proposed by Levinson
\& Waxman (2001). They have shown that, for typical microquasar
jet parameters, protons may be accelerated in the jet to
$\sim10^{16}$~eV, and that the interaction of these protons with
synchrotron photons emitted by the shock accelerated electrons is
expected to lead to 1--100~TeV neutrino emission. The predicted
fluxes are detectable by large, km$^2$-scale effective area,
high-energy neutrino telescopes, such as the operating south pole
detector AMANDA \cite{Andres00} and its planned 1~km$^2$
extension IceCube \cite{icecube}, or the Mediterranean sea
detectors under construction (ANTARES, see ANTARES Proposal;  NESTOR,
Monteleoni 1996) and planning (NEMO, Riccobene 2001; see
Halzen 2001 for a recent review).

In this paper we consider a class of identified Galactic
microquasars with either persistent jets or documented outbursts.
For each source we provide, for illustrative purposes, our model
prediction for the neutrino flux that should have been emitted
during particular events, using radio data available in the
literature. Although the temporal behaviour of many of these
sources may be unpredictable, we demonstrate that some of the
sources could have been detected by a neutrino telescope with
effective area larger than km$^2$ (in some cases even 0.1 km$^2$)
had such a detector been in operation during the time of the
recorded events and, therefore, propose that they should be
potential targets for the planned neutrino telescopes.  In
addition we consider a list of XRBs thus far unresolved at radio
wavelengths, that are believed to be microquasar candidates.
In \S\ref{sec:model} we briefly discuss the neutrino production
mechanism in microquasars. In \S\ref{sec:L_jet} we use
observational data available for each source to estimate the jet
parameters, and then use these parameters to derive the expected
neutrino flux. The number of neutrino induced muon events in
km$^2$-scale neutrino telescopes is derived in \S\ref{sec:N_mu}.
The implications of our results are briefly discussed in
\S\ref{sec:discussion}.

\section{Internal shock model for microquasars}
\label{sec:model}

In order to introduce the parameters relevant for the present
analysis, we give in this section a brief outline of the model
proposed by Levinson \& Waxman (2001) for production of neutrinos
in microquasars. The model assumes that on sufficiently small
scale ($\leq 10^{11}$ cm), inhomogeneities in the jet cause
internal shocks which can accelerate protons and electrons to a
power law distribution. The maximum proton energy is determined by
equating the acceleration time, estimated as the Larmor radius
divided by c, to the smallest of the dynamical time and the loss
time due to photomeson interactions. For typical jet parameters,
this energy is roughly $10^{16}$~eV in the jet frame. Protons can
interact both with the external X-ray photons emitted by the
accretion disc and with the synchrotron photons produced inside
the jet by the accelerated electrons, leading to pion production
and consequently to neutrino emission. In order for photomeson
production to take place, the comoving proton energy must exceed
the threshold energy for $\Delta$-resonance, $\approx10^{14}$ eV
for interaction with the external photons and $\approx10^{13}$ eV
for interaction with synchrotron photons.

Charged pions produced in photo-meson interactions decay to
produce neutrinos, $\pi^+ \rightarrow \mu^+\ +\ \nu_{\mu}
\rightarrow e^+ \ +\ \nu_e\ +\ \bar{\nu}_\mu+\ \nu_{\mu}$. In a
single collision, a pion is created with an average energy which
is $\approx20\%$ of the proton energy. This energy is roughly
evenly distributed between the final $\pi^+$ decay products,
yielding a $\nu$ energy which is $\approx5\%$ of the proton
energy.
The fraction of proton energy converted into pions, $f_{\pi}$,
depends on the jet Lorentz factor, $\Gamma$, and on  the kinetic
luminosity of the jet, $L_{jet}$. Levinson \& Waxman (2001) find
that for protons of energy
$\epsilon_p=\epsilon_{p,peak}\approx10^{13}$~eV, interacting with
the peak of the synchrotron photon spectrum, the fraction of
energy converted to pions is approximately given by
\begin{equation}
f_{\pi,peak}\simeq \min\left[1,0.1\eta_{e,-1}^{1/2}
\Gamma^{-2}\phi_{0.2}^{-1}L_{j38}^{1/2}\right],
\label{eq:fpi}
\end{equation}
where $L_{j}=10^{38}L_{j38}$ erg s$^{-1}$ and $\phi=0.2\phi_{0.2}$
are the jet kinetic power and opening angle, and
$\eta_{e}=0.1\eta_{e,-1}$ is the fraction of jet energy converted
to relativistic electrons (and hence to radiation). $f_\pi$
increases with proton energy; $f_\pi\propto\epsilon_p^{1/2}$
beyond the peak.
However, using the analysis of Levinson \& Waxman, we find that
the maximum proton energy for which the decay time of the $\pi^+$
produced in the photo-meson interaction is shorter than its
synchrotron energy loss time, is $\approx10\epsilon_{p,peak}$.
Since we are interested only in production of pions that decay to
produce neutrinos, and since the neutrino signal is dominated by
neutrinos in the energy range of 1 to 100~TeV, we use in what
follows the value given in Eq. (\ref{eq:fpi}) as the
characteristic value of $f_\pi$ for protons in the energy range
relevant for production of neutrinos that contribute to the
observed signal.

The expected fluence (energy per unit area) of $\gtsim1$~TeV muon
neutrinos at Earth from a jet ejection event is
\cite{Levinson01}
\begin{equation}
F_{\nu_\mu}\simeq \frac{1}{2}\eta_p
f_\pi\Gamma^{-1}\delta^3\frac{E_{jet}/8}{4\pi D^2},
\label{eq:fluence_nu}
\end{equation}
where $\delta=[\Gamma(1-\beta\cos\theta)]^{-1}$ is the Doppler
factor of the jet ($\theta$ is the angle between the jet axis and
our line of sight), $\eta_p\sim0.1$ the fraction of $L_{jet}$
carried by accelerated protons, and $D$ is the source distance.
$E_{jet}$ is the total energy carried by the jet during the event
duration (jet life time). The corresponding neutrino flux is
\begin{equation}
f_{\nu_\mu}\simeq \frac{1}{2}\eta_p
f_\pi\delta^4\frac{L_{jet}/8}{4\pi D^2}.
\label{eq:flux_nu}
\end{equation}

\section{Jet parameters and neutrino fluxes}
\label{sec:L_jet}

As explained in the introduction the term microquasar refers to
XRBs that reveal relativistic jets resolved in the radio band. In
events that have been monitored with sufficiently good resolution,
it is often possible to obtain a rough estimate of the
characteristic source parameters, in particular the bulk speed of
the jet, the angle between the jet axis and the sight line, and
the size of the emitting blob. Several other X-ray binaries, such
as Cygnus X-1, have been observed as point-like radio sources
\cite{Fender01b}. The spectrum  emitted by those sources and the
high degree of polarization sometimes measured is consistent with
synchrotron emission by nonthermal electrons.  Moreover, their
temporal behaviour appears to be similar to that seen in the
resolved microquasars and, therefore, it is tempting to conjecture
that they belong to the microquasar class. Since the putative jets
are unresolved in these objects, we use a different method to
estimate the kinetic power of their jets.  In what follows, we
consider separately the resolved and unresolved sources.

\subsection{Resolved microquasars}

In order to determine the neutrino flux from equation
(\ref{eq:flux_nu}), we use for each observed microquasar values of
$\Gamma$, $\delta$ and $D$ quoted in literature. $L_{jet}$ is
estimated in the following way: for a flux density
$S_\nu\propto\nu^{-1/2}$ of the source, implying an electron
energy distribution $dn_e/d\epsilon_e\propto\epsilon^{-2}$, the
minimum energy carried by electrons and magnetic field is obtained
for a magnetic field [e.g. \cite{Levinson96}]:
\begin{equation}
B_{*}=3.6\left[\ln(\gamma_{\rm max}/\gamma_{\rm min})
\frac{T_{B6}}{l_{15}} \right]^{2/7}
\frac{\nu^{5/7}_9}{\delta}\quad {\rm mG},
\label{eq:B_ep}
\end{equation}
where $\nu=\nu_9\cdot10^9$~Hz, $l=l_{15}\cdot10^{15}$~cm is the size of
the emission region, $\gamma_{\rm max}$ and $\gamma_{\rm min}$ are
the maximum and minimum electron random Lorentz factors as
measured in the jet frame, and $T_B=10^6T_{B6}$ K is the
brightness temperature: $T_B\equiv c^2 S_\nu/2\pi (l/D)^2 \nu^2$.
The minimum jet luminosity carried by electrons and magnetic field
is $L_{jet,eB}\ge (7/3) \pi l^2 c \Gamma^2 (B_*^2/8\pi)$. Denoting
by $\eta_e$ the fraction of jet kinetic energy converted to
internal energy of electrons and magnetic field, we finally obtain
\begin{equation}
L_{jet}\ge \frac{7}{24} c  \frac{(\Gamma l B_{*})^2}{ \eta_e }.
\label{eq:L_jet}
\end{equation}
For the numerical estimates that follow we conservatively assume
$\gamma_{\rm max}/\gamma_{\rm min}=100$.

As an example, consider the March 1994 event observed in GRS
1915+105.  This source is at a distance of $\sim 12.5$ kpc from
Earth. On March 24$^{th}$ 1994 the flux, measured with the VLA at
a wavelength of $3.5$ cm, was $655$ mJy.  Even though the source
was not resolved at the time, the inferred size of the blob was
$60\times 20$ mas \cite{Mirabel94}. Assuming a spherical blob
the corresponding blob radius is $l_{15}\sim2$. The estimated
speed of the ejecta is $\beta\sim0.92$ at an angle to the line of
sight of $\theta\sim70^\circ$. Using Eqs. (\ref{eq:L_jet}),
(\ref{eq:B_ep}) we obtain $T_B\sim4\cdot10^7~K$, $B_{*}\sim110$ mG
and $L_{jet}\sim2.5\cdot10^{40}\eta_{e,-1}^{-1}$ erg/sec. Here,
$\eta_e=0.1\eta_{e,-1}$.

A similar analysis was carried out for the other sources. In Table
\ref{tab:L_jet} we report our estimates of brightness temperature,
equipartition magnetic field and jet power for the list of known
microquasars, resolved in the radio band. In the same table we
also report the values of parameters required for our
calculations. The two values of the periodic source LS I
+61$^\circ$303 (P$\sim$26.5 days) refer respectively to bursting
and quiescent states observed by Massi et al. (2001).  The
parameters for V4641 Sgr are uncertain.  The distance to this
source has been estimated by Orosz et al. (2001) to lie in the
range between $7.4$ and $12.5$ kpc, based on their estimate of the
parameters of the companion star.  This is in contrast to a
distance of $\sim$ 0.5 kpc that seems to be favored by Hjellming
et al. (2000).  The former implies a jet with Lorentz factor
atypically high ($\Gamma\sim 22$), directed along our line of
sight and, consequently, a much higher neutrino flux.  In tables 1
and 2 we list the parameters and neutrino fluxes obtained for both
these distance estimates.

The steady source SS433 ($D=3$ kpc, $\beta\sim0.3$ and
$\theta\sim79^\circ$) is not present in Table \ref{tab:L_jet}. In
order to estimate the kinetic luminosity of the jet for this
source our simplified model cannot be applied. The source is
surrounded by the diffuse nebula W50, maybe a supernova remnant.
Several authors pointed out that the kinetic energy output of the
SS433 jets can influence the radio emission of W50; moreover SS433
is the only microquasar which shows a strong H$_\alpha$-line
emission from the jets \cite{Begelman80,Davidson80,Kirshner80}.
In our estimate we assume the conservative value of $L_{jet}\sim
10^{39}$ erg/sec suggested by Margon (1984).

In Table \ref{tab:flux1} we report the neutrino flux for the
listed microquasars, calculated using equation \ref{eq:flux_nu}
and considering the different values of $f_{\pi}$ given by Eq.
(\ref{eq:fpi}).

\subsection{Unresolved microquasars}

For the sources whose jet has not been resolved, we cannot deduce
the value of $L_{jet}$ from Eq.(\ref{eq:L_jet}), since $\Gamma$,
$\delta$ and the size of the jet are not known. We follow instead
a different line of argument. We define the jet synchrotron
luminosity as
\begin{equation}
L_{syn}= 4\pi D^2 \frac{1}{1-\alpha_R} S_{\nu_{high}} \nu_{high},
\label{eq:L_rad}
\end{equation}
where $\alpha_R$ is the spectral index, $\nu_{high}$ the highest
observed frequency of synchrotron emission, and  $S_{\nu_{high}}$
the flux density at this frequency.
We assume, as in the previous section, that a fraction $\eta_e\sim
0.1$ of the jet kinetic energy is converted to relativistic
electrons and magnetic field. The radio spectral index is
typically $\alpha_R\simeq 0.5$, implying that the electrons do not
cool fast on scales that are resolved by the VLA, and hence lose
only a small fraction of their energy to synchrotron emission. In
fact, assuming that the emission at $\nu_{high}$ originates from
the same radius as the radio emission, then for the typical
parameters inferred for the resolved MQs, that is, B on the order
of tens of mG and a corresponding synchrotron frequency of about
10 GHz, electrons radiating at $\nu_{high}\sim10^{15}$ Hz can not
lose more than $\eta_r\sim 0.1$ of their energy.  It could well be
however that the optical emission is produced at much smaller
radii, in which case $\eta_r$ may be of order unity. Given the
above, we estimate the jet luminosity as
\begin{equation}
L_{jet}= \eta_e^{-1}\eta_r^{-1} L_{syn},
\label{eq:Ljet2}
\end{equation}
and the neutrino flux at Earth as
\begin{equation}
f_{\nu_\mu}=\frac{1}{2}f_\pi \eta_p \frac{L_{jet}/8}{4\pi D^2}=
\frac{1}{16(1-\alpha_R)}\frac{\eta_p}{\eta_e} f_\pi \eta_r^{-1}
S_{\nu_{high}} \nu_{high}.
\label{eq:F_nu2}
\end{equation}
We have neglected in eqs. (\ref{eq:L_rad}--\ref{eq:F_nu2})
corrections due to relativistic expansion of the jets. However,
since these corrections are the same for both the synchrotron and
neutrino emission, the estimate of neutrino flux in eqs.
(\ref{eq:F_nu2}) is independent of such corrections.
In Table \ref{tab:L_jet2} we quote, for the sample of unresolved
microquasar candidates, the values of $\nu_{high}$ and
$S_{\nu_{high}}$, the distance of the source and our estimates of
$L_{jet}$ and $f_{\nu_\mu}$, calculated from Eq. (\ref{eq:Ljet2})
and Eq.(\ref{eq:F_nu2}).

\section{Muon events expected in a km$^2$-scale detector}
\label{sec:N_mu}

The detection of TeV neutrino fluxes from microquasars could be
the first achievable goal for proposed underwater(ice) neutrino
telescopes. In this section we calculate the rate of
neutrino-induced muon events expected in a detector with an
effective area of $1{\rm km}^2$. We also calculate the number of
atmospheric neutrino induced muon events expected in such a
detector in order to estimate the signal to noise ratio. Since the
signal is expected to be dominated by neutrinos of energy
$E_\nu\ge1$~TeV, for which the detection probability is
$P_{\nu\mu}\sim 1.3 \cdot 10^{-6} E_{\nu,TeV}$ \cite{Gaisser95},
we estimate the rate of neutrino-induced muon events as
\cite{Levinson01}:
\begin{equation}
\dot{N}_\mu \simeq 0.2 \eta_{p,-1}f_\pi\delta^4 D_{22}^{-2}
L_{jet,38} A_{{\rm km}^2} ~{\rm day}^{-1}
\label{eq:Mu_rate}
\end{equation}
where $A_{{\rm km}^2}$ is the effective detector area in km$^2$
units.
The total number of muon events in the detector is obtained by
multiplying the muon rate given by Eq. (\ref{eq:Mu_rate}) by the
duration, $\Delta t$, of the observed burst. In Table
\ref{tab:N_mu} we report the number of events expected in a
detector with $A_{eff}=1~{\rm km}^2$, during the bursts considered
in section \ref{sec:L_jet}. In the case of persistent sources, the
number of neutrino induced muon events, reported in Table
\ref{tab:N_mu}, is calculated for a 1 year period. In the same
table, we also report the number of atmospheric neutrino events
collected in such a detector during the time $\Delta t$. For the
background calculation, we assume a neutrino spectrum
$\phi_{\nu,bkg} \sim  10^{-7} E^{-2.5}_{\nu,\rm TeV}/ {\rm
cm}^2~{\rm sec}~{\rm sr}$ for $E_\nu>1$~TeV, $A_{{\rm km}^2}=1$
and a detector angular resolution of  0.3$^\circ$. Our
calculations suggest that microquasar signals may be identified
well above the atmospheric neutrino background by the next
generations of underwater(ice) neutrino telescopes.

In order to estimate the event rates for non-persistent sources it
is crucial to know their duty cycle. Some of these sources show a
periodic bursting activity: Circinus X-1 has a period of 16.59
days \cite{Preston83}, LS I +61 $^\circ$303 exhibits a 26.5 day
periodic non thermal outburst \cite{Massi01} and in the tables
we give the results for both the quiescent and the bursting phase.
Other transient sources show a stochastic bursting activity. For
such cases it is difficult to give an estimate of the expected
event rate. For example, in the campaign of August-December 1994
the lightcurves of GRO J1655-40 were dominated by three radio
flares, each lasting for 6 days \cite{Hjellming95}; In the
campaign of February-April 1994 GRS 1915+105 has exhibited 4
bursts \cite{Rodriguez99}.

\section{Discussion}
\label{sec:discussion}

There are large uncertainties involved in the derivation of the
jet parameters for most of the sources listed in table 4. The best
studied cases are perhaps GRS 1915+105, GRO J1655-40, and SS433.
Nonetheless, we have demonstrated that if the jets in microquasars
are protonic, and if a fraction of a few percent of the jet energy
is dissipated on sufficiently small scales, then emission of TeV
neutrinos with fluxes in excess of detection limit of the
forthcoming, km$^2$ scale, neutrino telescopes is anticipated.
Table \ref{tab:coordinates} shows the fraction of the time during
which each of the sources analyzed in this paper can be observed,
i.e. the number of hours per day during which the source is under
the horizon and may therefore produce upward moving muons, in a
northern hemisphere and a southern hemisphere detector.
The present identification of microquasars, and the inferred
distribution of their jet Lorentz factors, may be strongly
influenced by selection effects. It is quite likely that the class
of Galactic microquasars contains also sources with larger bulk
Lorentz factors and smaller viewing angles, which should emit
neutrinos with fluxes considerably larger than the extended
microquasars discussed in this paper.  Such sources may be
identified via their gamma-ray emission with, e.g. AGILE and GLAST, or by
their neutrino emission. The gamma-rays should originate from
larger scales where the pair production opacity is sufficiently
reduced. Predictions for AGILE and GLAST will be discussed 
in a forthcoming paper \cite{Guetta02}. There
are currently about 280 known XRBs \cite{Liu00,Liu01}, of
which $\sim 50$ are radio loud. These may also be potential
targets for the planned neutrino detectors.
Our results quoted in Tables \ref{tab:flux1} and \ref{tab:L_jet2}
are consistent with experimental upper limits on neutrino fluxes
from point sources set by MACRO \cite{Ambrosio01}, AMANDA
\cite{Biron} and SuperKamiokande \cite{Okada00}.
The MACRO upper limits \cite{Ambrosio01} for microquasar fluxes
are shown in table \ref{tab:macro}. For the brightest sources in
our table, SS433 and GX 339-4, the MACRO bounds are approximately
an order of magnitude larger than our predicted fluxes.

\acknowledgments
We particularly acknowledge E. Migneco for useful conversations.
We are grateful to R. Fender for stimulating 
discussions and useful informations.
D.G. and E.W. thank the director of the Laboratori Nazionali del Sud 
of Catania for the kind hospitality.

\clearpage

\begin{deluxetable}{lcccccccccccc}
\tabletypesize{\scriptsize}
\tablecaption{Brightness temperature, magnetic field and jet kinetic luminosity of identified Galactic microquasars, and the
parameters (inferred from observations) on which they are based (see text). \label{tab:L_jet}}
\tablewidth{0pt}
\tablehead{
\colhead{Source name}  &
\colhead{D}  &
\colhead{$\theta$}  &
\colhead{$\beta$}  &
\colhead{$\Gamma$}  &
\colhead{$\delta$}  &
\colhead{l$_{15}$}  &
\colhead{$\nu$}   &
\colhead{$S_\nu$ }   &
\colhead{$T_B$}   &
\colhead{$B_{*}$}   &
\colhead{ $\eta_{e,-1}L_{jet}$}   &
\colhead{Ref.}   \\
\colhead{}  &
\colhead{(kpc)}  &
\colhead{(deg)}  &
\colhead{}  &
\colhead{}  &
\colhead{}  &
\colhead{}  &
\colhead{(GHz)}   &
\colhead{ (mJy)}   &
\colhead{(K)}   &
\colhead{(mG)}   &
\colhead{ (erg/sec)}   &
\colhead{}
}
\startdata
CI Cam              & 1    & 83 & 0.15  & 1.01 & 1.01 & 0.8  & 22.5 & 450   & 1.57$\cdot10^{5}$ & 33.5 & 5.66$\cdot10^{37}$ & 1,2  \\
XTE J1748-288       & 8    & 64 & 0.73  & 1.46 & 1.01 & 15.0 & 4.9  & 410   & 4.90$\cdot10^{5}$ & 6.6  & 1.84$\cdot10^{39}$ & 3,4,5    \\
Cygnus X-3          & 10   & 14 & 0.81  & 1.70 & 2.74 & 0.23 & 15   & 10500 & 8.47$\cdot10^{9}$ & 288.7& 1.17$\cdot10^{39}$ & 6  \\
LS 5039             & 3    & 68 & 0.4   & 1.09 & 1.08 & 0.04 & 5.0  & 16    & 2.83$\cdot10^{8}$ & 203  & 8.73$\cdot10^{36}$ & 7  \\
GRO J1655-40        & 3.1  & 81 & 0.92  & 2.55 & 0.46 & 4.7  & 1.6  & 3400  & 5.86$\cdot10^{7}$ & 35.9 & 1.60$\cdot10^{40}$ & 8    \\
GRS 1915+105        & 12.5 & 70 & 0.92  & 2.55 & 0.57 & 1.9  & 8.6  & 655   & 3.94$\cdot10^{7}$ & 110  & 2.45$\cdot10^{40}$ & 9    \\
Circinus X-1        & 10   & 70 & 0.1   & 1.01 & 1.03 & 1.1  & 2.3  & 1200  & 1.78$\cdot10^{9}$ & 82.3 & 7.61$\cdot10^{38}$ & 4,10   \\
LS I 61$^\circ$303  & 2    &0.2 & 0.43  & 1.11 & 1.58 & 0.2  & 5.0  & 220   & 1.55$\cdot10^{8}$ & 82.6 & 1.65$\cdot10^{37}$ & 11    \\
LS I 61$^\circ$303  & 2    &0.2 & 0.43  & 1.11 & 1.58 & 0.2  & 5.0  & 34    & 2.40$\cdot10^{7}$ & 48.4 & 5.69$\cdot10^{36}$ & 11 \\
XTE J1550-564       & 2.5  & 74 & 0.83  & 1.79 & 0.72 & 1.1  & 2.3  & 71    & 6.64$\cdot10^{6}$ & 23.7 & 2.01$\cdot10^{38}$ & 12,13  \\
V4641 Sgr           & 0.5  & 63 & 0.85  & 1.90 & 0.86 & 0.9  & 4.9  & 400   & 4.71$\cdot10^{5}$ & 17.0 & 8.02$\cdot10^{37}$ & 14     \\
V4641 Sgr           & 9.6  & 6 & 0.999 & 22.37 & 6.91 & 18.0 & 4.9  & 400   & 4.71$\cdot10^{5}$ & 0.91 & 1.17$\cdot10^{40}$ & 14,15  \\
Scorpius X-1	    & 2.8  & 44 & 0.95  & 3.20 & 0.99 & 0.06 & 5    & 20    & 1.57$\cdot10^{8}$ &170.4 & 1.04$\cdot10^{38}$ & 16,17  \\
\enddata
\tablerefs{
(1) Hjellming \& Mioduszewski 1998, (2) Harmon \& Fishman 1998, (3)
Rupen \& Hjellming 1998, (4) Mirabel \& Rodr\'{\i}guez 1999, (5) Kotani {\it et al.} 2000, (6) Mioduszewski {\it et
al.} 2001, (7) Paredes {\it et al.} 2000, (8) Hjellming \& Rupen 1995, (9) Mirabel \& Rodr\'{\i}guez 1994, (10)
Preston {\it et al.} 1983, (11) Massi {\it et al.} 2001, (12) Hannikainen {\it et al.} 2001, (13) Sanches-Fernandez
1999, (14) Hjellming {\it et al.} 2000, (15) Orosz {\it et al.} 2001, (16) Fomalont {\it et al.} 2001a, (17)
Fomalont {\it et al.} 2001b.} 
\end{deluxetable}

\clearpage

\begin{table}
\caption{$>1$~TeV neutrino flux at Earth from identified
Galactic microquasars, estimated using Eq. (\ref{eq:flux_nu}). \label{tab:flux1}}
\begin{center}
\begin{small}
\begin{tabular}{lcc}
\tableline \tableline
\multicolumn{3}{}{}\\
Source name         & $f_{\pi,peak}$ & $\eta_{p,-1}^{-1}\eta_{e,-1}^{1/2}f_{\nu}$ (erg/ cm$^2$ sec) \\
\tableline
CI Cam                  & 0.07  & 2.23$\cdot10^{-10}$ \\
XTE J1748-288           & 0.20  & 3.07$\cdot10^{-10}$ \\
Cygnus X-3              & 0.12  & 4.02$\cdot10^{-9}$  \\
LS 5039                 & 0.02  & 1.69$\cdot10^{-12}$ \\
GRO J1655-40            & 0.19  & 7.37$\cdot10^{-10}$ \\
GRS 1915+105            & 0.24  & 2.10$\cdot10^{-10}$ \\
Circinus X-1            & 0.27  & 1.22$\cdot10^{-10}$ \\
LS I 61$^\circ$303      & 0.03  & 4.49$\cdot10^{-11}$ \\
LS I 61$^\circ$303      & 0.02  & 9.06$\cdot10^{-12}$ \\
XTE J1550-564           & 0.04  & 2.00$\cdot10^{-11}$ \\
V4641 Sgr               & 0.02  & 2.25$\cdot10^{-10}$ \\
V4641 Sgr               & 0.002 & 3.25$\cdot10^{-8}$ \\
Scorpius X-1	        & 0.01  & 6.48$\cdot10^{-12}$ \\
SS433                   & 0.29  & 1.72$\cdot10^{-9}$  \\
\tableline \tableline
\end{tabular}
\end{small}
\end{center}
\end{table}

\clearpage

\begin{table}
\caption{Kinetic jet luminosity, $L_{jet}$, and $>1$~TeV
neutrino flux at Earth, $f_{\nu_\mu}$, for unresolved
microquasars, estimated using Eqs. (\ref{eq:Ljet2}) and
(\ref{eq:F_nu2}). D is the source-observer distance, $\nu_{high}$
is the highest frequency at which synchrotron emission is
observed, $S_{\nu}$ is the flux density measured at $\nu =
\nu_{high}$. \label{tab:L_jet2}} 
\begin{center}
\begin{small}
\begin{tabular}{lcccccc}
\tableline \tableline
\multicolumn{7}{}{}\\
Source name & D  & $\nu_{high}$& $S_{\nu}$ & $L_{jet}$& $(\eta_e/\eta_p)(\eta_{r,-1}/f_\pi)f_{\nu_\mu}$  & Ref.\\
                        &(kpc)  & ($10^{14}$ Hz)  & (mJy)     & (erg/sec)        & (erg/cm$^2$ sec)  & \\
\tableline
GS 1354-64      & 10 & 3  & 5   & 3.62$\cdot10^{37}$ & 1.88$\cdot10^{-11}$ & 1,2\\
GX 339-4        & 4  & 10 & 100 & 3.86$\cdot10^{38}$ & 1.26$\cdot10^{-9}$  & 1,3,4\\
Cygnus X-1      & 2  & 1  & 15  & 1.45$\cdot10^{36}$ & 1.88$\cdot10^{-11}$ & 3,5\\
GRO J0422+32    & 3  & 4  & 50  & 4.35$\cdot10^{37}$ & 2.51$\cdot10^{-10}$ & 1,6\\
XTE J1118+480   &1.9 & 20 & 20  & 3.49$\cdot10^{37}$ & 5.02$\cdot10^{-10}$ & 1,7\\
\tableline \tableline
\end{tabular}
\end{small}
\end{center}
\tablerefs{
(1) Fender 2000, (2) Brocksopp 2001, (3) Fender 2001b, (4) Corbel {\it et al.} 2000, (5) Pooley {\it et al.} 1999, (6) Shrader 1994, (7) Frontiera 2001. }
\end{table}

\clearpage

\begin{table}
\caption{Predicted number of muon events, N$_\mu$, in a
$1{\rm km}^2$ detector, based on the neutrino fluxes given in
tables \ref{tab:flux1} and \ref{tab:L_jet2}. For bursting sources,
the number of events expected in a burst of duration $\Delta t$ is
quoted. For persistent sources, 1 year integration time is
assumed. The two values quoted for the periodic source LS I
+61$^\circ$303 refer to a bursting state of duration 7 days and to
a quiescent state of duration $\sim$ 20 days. N$_{\mu,bkg}$ is the
expected number of background atmospheric neutrino events, assuming
angular resolution of 0.3$^\circ$. \label{tab:N_mu}} 
\begin{center}
\begin{small}
\begin{tabular}{lccc}
\tableline \tableline
\multicolumn{4}{}{}\\
Source name & $\Delta t$ (days) & N$_\mu$& N$_{\mu,bkg}(deg/0.3^\circ)^2$      \\
\tableline
CI Cam                  & 0.6  & 0.05 & 0.002    \\
XTE J1748-288           & 20   & 2.5 & 0.054     \\
Cygnus X-3              & 3    & 4.8 & 0.008     \\
LS 5039                 & persistent & 0.2 & 0.986  \\
GRO J1655-40            & 6    & 1.8 & 0.016     \\
GRS 1915+105            & 6    & 0.5 & 0.016     \\
Circinus X-1            & 4    & 0.2 & 0.011     \\
LS I 61$^\circ$303      & 7    & 0.1 & 0.019     \\
LS I 61$^\circ$303      & 20   & 0.1 & 0.054     \\
XTE J1550-564           & 5    & 0.04 & 0.014    \\
V4641 Sgr               & 0.3  & 0.03 & 0.001    \\
V4641 Sgr               & 0.3  & 3.9 & 0.001     \\
Scorpius X-1		& persistent & 0.9 & 0.986 \\
SS433                   & persistent & 252 & 0.986 \\
GS 1354-64              & 2.8  & 0.02 & 0.008      \\
GX 339-4                & persistent & 183.4 & 0.986 \\
Cygnus X-1              & persistent & 2.8 & 0.986  \\
GRO J0422+32            & 1$\div$20   & 0.1$\div$2 & 0.003$\div$0.1 \\
XTE J1118+480           & 30$\div$150 & 6$\div$30 & 0.081$\div$0.405   \\
\tableline \tableline
\end{tabular}
\end{small}
\end{center}
\end{table}

\clearpage

\begin{table}
\caption{Microquasar equatorial coordinates and the
fraction of time (hours per day) during which the source is below
the horizon, and can hence be observed by a neutrino telescope,
for a telescope located at Lat. 36$^\circ$ 23' 51" N (NEMO,
Mediterranean sea), $T_N$, and for a neutrino telescope located at
the South Pole (Amanda-IceCube), $T_I$.\label{tab:coordinates}
} \begin{center}
\begin{small}
\begin{tabular}{lcclcc}
\tableline \tableline
\multicolumn{6}{}{}\\
Source name             & Right Ascension        & Declination & & $T_N$ & $T_I$   \\
                         & & & & (hours) & (hours)  \\
\tableline
 CI Cam                  & ~4$^h$ 19$^m$ 42$^s$.2 & +55$^\circ$ 59' 58".0 & (2000)       & ~0    & 24     \\
 XTE J1748-288           & 17$^h$ 48$^m$ 05$^s$.1 & -28$^\circ$ 28' 25".8 & (2000)       & 15.1   & 0  \\
 Cygnus X-3              & 20$^h$ 32$^m$ 26$^s$.6 & +40$^\circ$ 57' 09".0 & (2000)       & ~6.7   & 24     \\
 LS 5039                 & 18$^h$ 26$^m$ 14$^s$.9 & -14$^\circ$ 50' 51".0 & (2000)       & 13.5   & 0     \\
 GRO J1655-40            & 16$^h$ 54$^m$ ~0$^s$.2 & -39$^\circ$ 50' 44".7 & (2000)       & 17.0   & 0     \\
 GRS 1915+105            & 19$^h$ 12$^m$ 50$^s$.0 & +10$^\circ$ 51' 27".0 & (1950)       & 10.9   & 24     \\
 SS433                   & 19$^h$ 11$^m$ 49$^s$.6 & +04$^\circ$ 58' 58".0 & (2000)       & 11.5   & 24     \\
 Circinus X-1            & 15$^h$ 16$^m$ 48$^s$.3 & -56$^\circ$ 59' 14".0 & (1950)       & 24     & 0       \\
 LS I +61$^\circ$303     & ~2$^h$ 40$^m$ 31$^s$.7 & +61$^\circ$ 13' 45".6 & (2000)       & ~0     & 24 \\
 XTE J1550-564           & 15$^h$ 50$^m$ 58$^s$.7 & -56$^\circ$ 28' 35".3 & (2000)       & 24     & 0 \\
 V4641 Sgr               & 18$^h$ 19$^m$ 21$^s$.5 & -25$^\circ$ 25' 36".0 & (2000)       & 14.7   & 0 \\
 Scorpius X-1		 & 16$^h$ 19$^m$ 55$^s$.1 & -15$^\circ$ 38' 24".9 & (2000)       & 13.6   & 0  \\
 GS 1354-64              & 13$^h$ 58$^m$ 09$^s$.7 & -64$^\circ$ 44' 05".0 & (2000)       & 24     & 0 \\
 GX 339-4                & 17$^h$ 02$^m$ 49$^s$.5 & -48$^\circ$ 47' 23".0 & (2000)       & 19.6   & 0 \\
 Cygnus X-1              & 19$^h$ 58$^m$ 21$^s$.7 & +35$^\circ$ 12' 05".8 & (2000)       & ~7.8   & 24 \\
 GRO J0422+32            & ~4$^h$ 21$^m$ 42$^s$.7 & +32$^\circ$ 54' 27".0 & (2000)       & ~8.2   & 24 \\
 XTE J1118+480           & 11$^h$ 18$^m$ 17$^s$.0 & +48$^\circ$ ~3' ~0".0 & (2000)       & ~4.7   & 24     \\
 \tableline \tableline
 \end{tabular}
 \end{small}
 \end{center}
 \end{table}

\clearpage

\begin{table}
\caption{Upper limit (90\% c.l.) on $>1$~TeV neutrino
fluxes from individual miqroquasars, inferred from the MACRO upper
limit on the number flux of $>1$~GeV neutrinos, $f_\#(>1{\rm
GeV})$, from these sources (\cite{Ambrosio01}). Since the MACRO
limit is derived assuming a neutrino spectrum
$dn_\nu/d\epsilon_\nu\propto\epsilon^{-2}$, the inferred limit on
1~TeV to 100~TeV neutrino energy flux is
$\approx\ln(100)f_\#(>1{\rm GeV})\times1$~GeV. \label{tab:macro}
} \begin{center}
\begin{small}
\begin{tabular}{lc}
\tableline \tableline
\multicolumn{2}{}{}\\
Source name             & MACRO upper limits (erg cm$^{-2}$ s$^{-1}$) \\
\tableline
Circinus X-1      & $1.77\times 10^{-8}$ \\
GX 339-4          & $2.40\times 10^{-8}$ \\
SS433             & $1.02\times 10^{-8}$ \\
Cygnus X-1        & $4.99\times 10^{-8}$ \\
Cygnus X-3        & $9.99\times 10^{-8}$ \\
Scorpius X-1      & $1.27\times 10^{-8}$ \\
\tableline \tableline
\end{tabular}
\end{small}
\end{center}
\end{table}

\end{document}